\documentstyle{merritt}

\newif\ifAMStwofonts
\def\fun#1#2{\lower3.6pt\vbox{\baselineskip0pt\lineskip.9pt
  \ialign{$\mathsurround=0pt#1\hfil##\hfil$\crcr#2\crcr\sim\crcr}}}
\def\lap{\mathrel{\mathpalette\fun <}}
\def\gap{\mathrel{\mathpalette\fun >}}

\ifoldfss
  \ifCUPmtlplainloaded \else
    \NewTextAlphabet{textbfit} {cmbxti10} {}
    \NewTextAlphabet{textbfss} {cmssbx10} {}
    \NewMathAlphabet{mathbfit} {cmbxti10} {} % for math mode
    \NewMathAlphabet{mathbfss} {cmssbx10} {} %  "   "    "
  \fi
  \ifAMStwofonts
    \ifCUPmtlplainloaded \else
      \NewSymbolFont{upmath} {eurm10}
      \NewSymbolFont{AMSa} {msam10}
      \NewMathSymbol{\upi}     {0}{upmath}{19}
      \NewMathSymbol{\umu}     {0}{upmath}{16}
      \NewMathSymbol{\upartial}{0}{upmath}{40}
      \NewMathSymbol{\leqslant}{3}{AMSa}{36}
      \NewMathSymbol{\geqslant}{3}{AMSa}{3E}

       \let\le=\leqslant
       \let\ge=\geqslant
    \fi
  \fi
\fi % End of OFSS

\ifnfssone
  \newmathalphabet{\mathit}
  \addtoversion{normal}{\mathit}{cmr}{m}{it}
  \addtoversion{bold}{\mathit}{cmr}{bx}{it}
  \newmathalphabet{\mathbfit} % math mode version of \textbfit{..}
  \addtoversion{normal}{\mathbfit}{cmr}{bx}{it}
  \addtoversion{bold}{\mathbfit}{cmr}{bx}{it}
  \newmathalphabet{\mathbfss} % math mode version of \textbfss{..}
  \addtoversion{normal}{\mathbfss}{cmss}{bx}{n}
  \addtoversion{bold}{\mathbfss}{cmss}{bx}{n}
  \ifAMStwofonts
    \ifCUPmtlplainloaded \else
      \UseAMStwoboldmath
      \makeatletter
      \new@mathgroup\upmath@group
      \define@mathgroup\mv@normal\upmath@group{eur}{m}{n}
      \define@mathgroup\mv@bold\upmath@group{eur}{b}{n}
      \edef\UPM{\hexnumber\upmath@group}
      \new@mathgroup\amsa@group
      \define@mathgroup\mv@normal\amsa@group{msa}{m}{n}
      \define@mathgroup\mv@bold\amsa@group{msa}{m}{n}
      \edef\AMSa{\hexnumber\amsa@group}
      \makeatother
      \mathchardef\upi="0\UPM19
      \mathchardef\umu="0\UPM16
      \mathchardef\upartial="0\UPM40
      \mathchardef\leqslant="3\AMSa36
      \mathchardef\geqslant="3\AMSa3E

       \let\le=\leqslant
       \let\ge=\geqslant
    \fi
  \fi
\fi % End of NFSS release 1

\ifnfsstwo
  \DeclareMathAlphabet{\mathbfit}{OT1}{cmr}{bx}{it}
  \SetMathAlphabet\mathbfit{bold}{OT1}{cmr}{bx}{it}
  \DeclareMathAlphabet{\mathbfss}{OT1}{cmss}{bx}{n}
  \SetMathAlphabet\mathbfss{bold}{OT1}{cmss}{bx}{n}
  \ifAMStwofonts
    \ifCUPmtlplainloaded \else
      \DeclareSymbolFont{UPM}{U}{eur}{m}{n}
      \SetSymbolFont{UPM}{bold}{U}{eur}{b}{n}
      \DeclareSymbolFont{AMSa}{U}{msa}{m}{n}
      \DeclareMathSymbol{\upi}{0}{UPM}{"19}
      \DeclareMathSymbol{\umu}{0}{UPM}{"16}
      \DeclareMathSymbol{\upartial}{0}{UPM}{"40}
      \DeclareMathSymbol{\leqslant}{3}{AMSa}{"36}
      \DeclareMathSymbol{\geqslant}{3}{AMSa}{"3E}

       \let\le=\leqslant
       \let\ge=\geqslant
    \fi
  \fi
\fi % End of NFSS release 2

\ifCUPmtlplainloaded \else
  \ifAMStwofonts \else % If no AMS fonts
    \def\upi{\pi}
    \def\umu{\mu}
    \def\upartial{\partial}
  \fi
\fi
\title{Orbital instability and relaxation in stellar systems}

\author[Monica Valluri and David Merritt]
       {Monica Valluri$^{1,2}$ and David Merritt$^1$\\
$^1$ Department of Physics and Astronomy, Rutgers University,
New Brunswick, NJ 08855.\\
$^2$ University of Chicago, Astronomy \& Astrophysics Center, 5640 S. Ellis Ave, Chicago IL 60637.\\
e-mail:valluri@oddjob.uchicago.edu, merritt@astro.rutgers.edu}
\date{Rutgers Astrophysics Preprint Series No. 257}

\begin{document}

\maketitle

\label{firstpage}

\begin{abstract}
We review recent progress in understanding the role of chaos in
influencing the structure and evolution of galaxies.  The orbits of
stars in galaxies are generically chaotic: the chaotic behavior arises
in part from the intrinsically grainy nature of a potential that is
composed of point masses.  Even if the potential is assumed to be
smooth, however, much of the phase space of non-axisymmetric galaxies
is chaotic due to the presence of central density cusps or black
holes.  The chaotic nature of orbits implies that perturbations will
grow exponentially and this in turn is expected to result in a
diffusion in phase space. We show that the degree of orbital evolution
is not well predicted by the growth rate of infinitesimal
perturbations, i.e. by the Liapunov exponent.  A more useful criterion
is whether perturbations continue to grow exponentially until their
scale is of order the size of the system.  We illustrate these ideas
in a potential consisting of $N$ fixed point masses. Liapunov
exponents are large for all values of $N$, but orbits become
increasingly regular in their behavior as $N$ increases; the reason is
that the exponential divergence saturates at smaller and smaller
distances as $N$ is increased.  The objects which lend phase space its
structure and impede diffusion are the invariant tori. In the triaxial
potentials we discuss, a large fraction of the tori correspond to
resonant (thin) orbits and their associated families of regular
orbits. These tori are destroyed by perturbations to the
potential. When only a few stable resonances remain, we find that the
phase space distribution of an ensemble of chaotic orbits evolves
rapidly toward a nearly stationary state.  This mixing process is
shown to occur on timescales of a few crossing times in triaxial
potentials containing massive central singularities, consistent with
the rapid evolution observed in $N$-body simulations of galaxies with
central black holes.
\end{abstract}

\section{Introduction}

The role of microscopic chaos in producing macroscopic relaxation of
dynamical systems has received considerable attention in the field of
statistical mechanics (\cite{le94}, \cite{gas98}).  
However, the role of chaos in the relaxation of stellar systems is 
relatively less well understood.  
The last five years have witnessed the development of a number of
new techniques for probing the complexities of phase space in realistic 
galactic potentials. 
This work has led to a greater understanding of the importance of stochastic
orbits and their role in driving dynamical evolution.

The gravitational forces on a star in a galaxy can be broken up into
two components: a rapidly varying component that arises from the
discrete distribution of stars, and a smoothly varying component that
arises from the overall mass distribution.  
The importance of the discrete component of the force relative to the 
smooth component is usually assumed proportional to $\sim 10\ln{N}/N$, 
the ratio of dynamical to two-body relaxation times (\cite{ch43}).
The implication is that, for galaxies ($N \sim 10^{11}$), the time
scale on which the discrete component of the force is important
is much longer than the age of the Universe. 
At the same time, it has been shown (\cite{mi64}, \cite{gs86}) that 
trajectories in the $N$-body problem are generically chaotic, 
with rates of exponential divergence that appear to remain large even
for large $N$. Thus, according to at least one definition of chaos, 
the orbits of stars in galaxies should never tend toward the regular 
motion expected in smooth potentials.

The exponential divergence of trajectories gives rise to a diffusion.
The approach of a non-stationary distribution of phase points to a 
stationary one via chaotic diffusion is referred to as ``chaotic mixing'' 
or just ``mixing'' (\cite{km94}). 
In the context of non-equilibrium statistical mechanics, 
mixing to an invariant distribution is regarded as a legitimate relaxation
process (\cite{le94}, \cite{gas98}). However, in the context of gravitational systems, the issue of
whether or not there is a connection between the time scale for
exponential divergence of adjacent trajectories and the
time scale for macroscopic relaxation has been a matter of much debate
(\cite{gs86}, \cite{kan90}, \cite{ks91}, \cite{heg91}, \cite{ghh93}).

The question of how and under what conditions chaos will result
in evolution of a stellar system is the subject of this article.
In \S2 we present two idealized models: 
a set of $N$ fixed mass points, representing stars in a galaxy;
and a smooth potential with a single central point mass, 
representing a galaxy with a nuclear black hole.
Motion in both potentials is generically chaotic, with similar 
values of the Liapunov exponent.
We show that the Liapunov exponent -- which measures only the 
infinitesimal growth of perturbations -- is not always useful 
for predicting the degree of macroscopic evolution.
A more useful criterion is whether perturbations continue to grow 
exponentially until their size is of order the size of the system.
The kinds of structures that can impede diffusion of stochastic orbits 
are discussed in \S3.
In \S4 we simulate the approach to an invariant distribution of
stars in galactic potentials; we find that evolution to a 
stationary state can take place in little more than a crossing time 
if the phase space is globally chaotic.
Such evolution is consistent with the rapid, self-consistent 
evolution observed in numerical simulations of galaxies 
containing nuclear black holes.

\section{The Liapunov exponent and other measures of chaos}

Exponential divergence of nearby trajectories is a common property of
dynamical systems.  This divergence is most commonly expressed in
terms of the Liapunov exponent, which measures the mean $e$-folding
rate of an infinitesimal perturbation averaged over an infinite time
interval.  Because the perturbation is assumed to remain small,
Liapunov exponents measure only the rate of divergence in the
immediate vicinity of the unperturbed orbit; they do not necessarily
contain any information about the non-linear, or macroscopic, evolution.  
One might nevertheless expect the magnitude of the Liapunov
exponent to predict, in some approximate sense, the degree to which
chaos will induce finite changes in the structure of an orbit
after a fixed time.  
This turns out not always to be the case, as we now illustrate.

Figures 1 and 2 summarize the results of test-particle integrations in
two time-independent potentials.  Model 1 (Fig. 1) consists of $N$
point masses $m$ distributed randomly and uniformly within a triaxial
ellipsoid; the total mass $M=Nm$ remains fixed ($M=1$) as $N$ is varied,
as do the axis lengths ($a=1, b=0.75, c=0.5$) of the ellipsoid.  
Model 2 (Fig. 2) is
the smooth representation of Model 1, i.e. a homogeneous ellipsoid, to
which has been added a single central point of mass $M_h$.  The
parameter $N$ of Model 1 may be interpreted as the number of stars
that make up a galaxy of fixed mass, and the parameter $M_h$ of Model
2 as the mass of a central black hole, expressed in units of the total
galaxy mass.  The quantities $1/N$ and $M_h$ play the role of
perturbation parameters; as they are increased, the potential departs
more and more from that of the integrable, 3D harmonic oscillator, and
one expects to see corresponding changes in the behavior of orbits.

\begin{figure*}
\vspace{13.cm}
\includegraphics{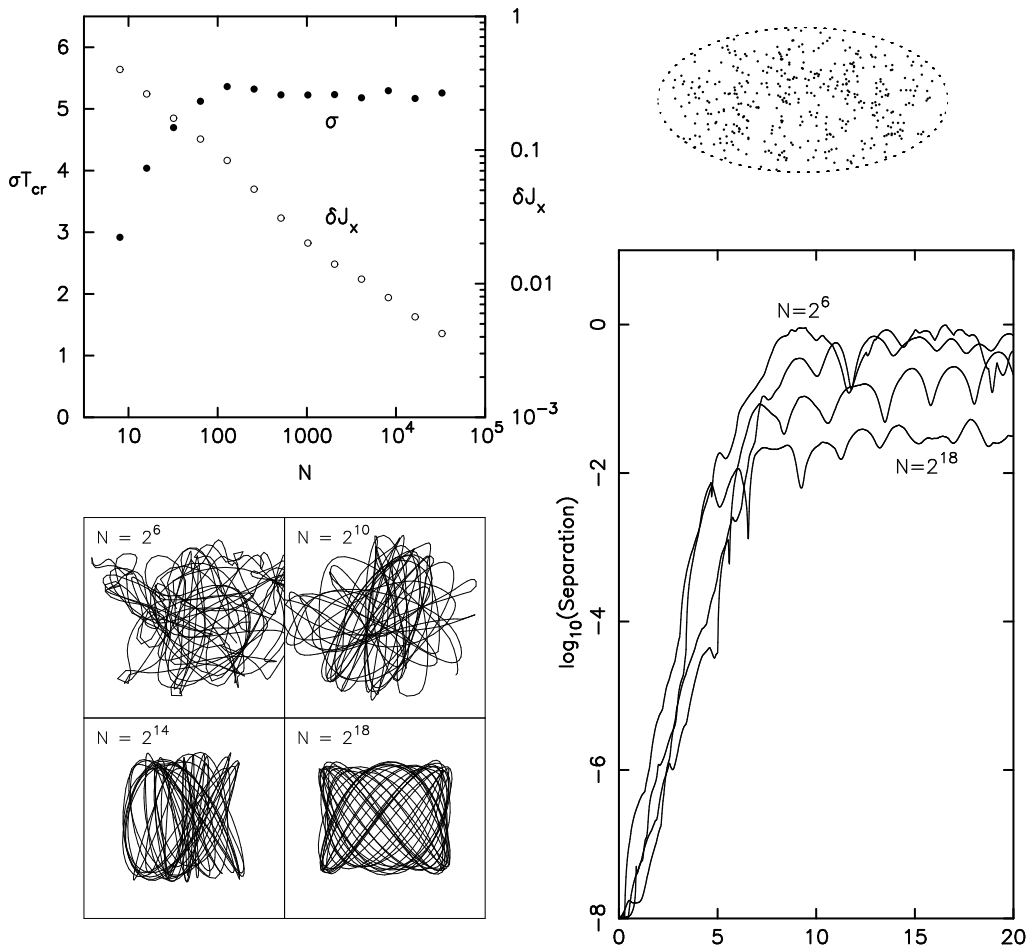}
\caption{ Evolution of orbits in a potential consisting of $N$ fixed
point masses, $m=1/N$, distributed randomly and uniformly in an
ellipsoidal volume.  For each $N$, 32 random realizations of the
particle positions were generated and a single orbit was integrated in
each of the corresponding potentials.  The upper left panel shows mean
values of the Liapunov exponent $\sigma$ for each ensemble, and the
RMS variation in the ``action'' $J_x$ over $\sim 20$ orbital periods
(see text).  $\sigma$ remains large while $\delta J_x$ falls off with
$N$; thus the behavior of a typical orbit becomes more and more
regular as $N$ increases.  This is verified in the lower left panel
which shows typical orbits for four values of $N$.  The lower right
panel displays the {\it finite} separation between initially adjacent
trajectories.  While the divergence takes place at the rate predicted
by the Liapunov exponents at early times, for large $N$ the separation
saturates at a value much smaller than the size of the system.
\label{fig:Nbdy}}	
\end{figure*}

The evolution of orbits in Model 1 is illustrated in Fig. 1, as a
function of the number of particles making up the potential.  For each
$N$, a set of 32 random realizations of the particle positions were
generated; a single orbit was then integrated in each of the 32
corresponding potentials.  The evolution of the orbits were measured
in several ways.  The mean Liapunov exponent $\sigma$ (averaged over
the 32 ensembles) was found to increase with $N$ until $N\approx 100$
then level off, at a value $\sigma T_{cr}\sim 5$, with $T_{cr}$
defined as one-half of the long-axis orbital period in the smooth
potential.  Thus these trajectories are locally unstable on a
remarkably short time scale, just a fraction of a crossing time, and
this characteristic time shows no tendency to decrease with increasing
$N$ for $N$ as large as $\sim 10^5$.  This result is similar to the
well-known exponential instability of the full $N$-body problem
(\cite{mi64}, \cite{gs86}, \cite{kan90}, \cite{ks91}, \cite{heg91},
\cite{ghh93}).

As a second measure of the orbital evolution, the RMS variations in
the $x$, $y$ and $z$ ``actions'' of the orbits were computed over
$\sim 20$ periods; these ``actions'' were defined in terms of the
frequencies of motion in the smooth potential,
e.g. $J_x=E_x/\omega_x=(v_x^2+\omega_x x^2)/(2\omega_x)$, and would be
precisely conserved in the limit of zero perturbation.  Figure 1 shows
that there is a smooth decrease in the amplitude of $\delta J_x$ as
$N$ is increased -- in other words, the orbits approach more and more
closely, in their {\it macroscopic} behavior, to that of the
integrable orbits even though they remain {\it locally} unstable to a
degree (as measured by $\sigma$) that is nearly independent of $N$.
Plots of the trajectories (Fig. 1) confirm that the orbits for large
$N$ are similar to the Lissajous figures expected in the 3D harmonic
oscillator.

This apparent paradox -- large $\sigma$, but nearly regular 
motion -- is reconciled in the lower right panel of Fig. 1,
which shows the variation with time of the {\it finite} separation
between two initially nearby trajectories, for four values of $N$.  
The divergence is initially exponential in all cases, with a
characteristic time $\sim 0.2T_{cr}$, consistent with the 
measured values of $\sigma$.  
However the exponential divergence eventually saturates, after which 
the separation slows.
This saturation has no effect on the Liapunov exponents which 
are calculated assuming that the separation remains infinitesimally 
small. Furthermore, the saturation occurs sooner for larger $N$.  
When $N\lap 10^3$, the exponential divergence
continues until the separation is of order the size of the system,
while for $N\approx 10^5$, saturation occurs at a separation of only
$\sim 10^{-2}$, much smaller than the system size.  

How can the separation between two, locally unstable orbits saturate
at a small amplitude, given that neither orbit ``knows''
about the other orbit?  The answer must be that both
orbits are confined to the same, restricted region of
phase space; saturation occurs when the separation between them is of
order the width of this region.  
Further systematic increase in
their separation would require that one of the trajectories ``break
out'' into another such region, and such events (for large
$N$) apparently occur at a much lower rate than the divergence
described by the Liapunov exponent.  The fact that the exponential
divergence saturates sooner for larger $N$ suggests that the width of
the confining regions decreases with increasing $N$ -- although our
experiments do not allow us to estimate the precise $N$-dependence.
The apparently unconfined evolution for $N\lap 10^3$ suggests that
phase space for such small $N$ is ``globally chaotic'', with 
essentially no confining barriers.

Figure 1  suggests the way in which the equations of motion in 
an $N$-body potential tend toward those of the corresponding smooth 
potential as $N$ increases.
Although some measures of chaos -- e.g. the Liapunov exponents -- 
remain large even for large $N$, others -- e.g. the 
change in the actions -- tend to zero.

\begin{figure*}
\vspace{13cm}
\includegraphics{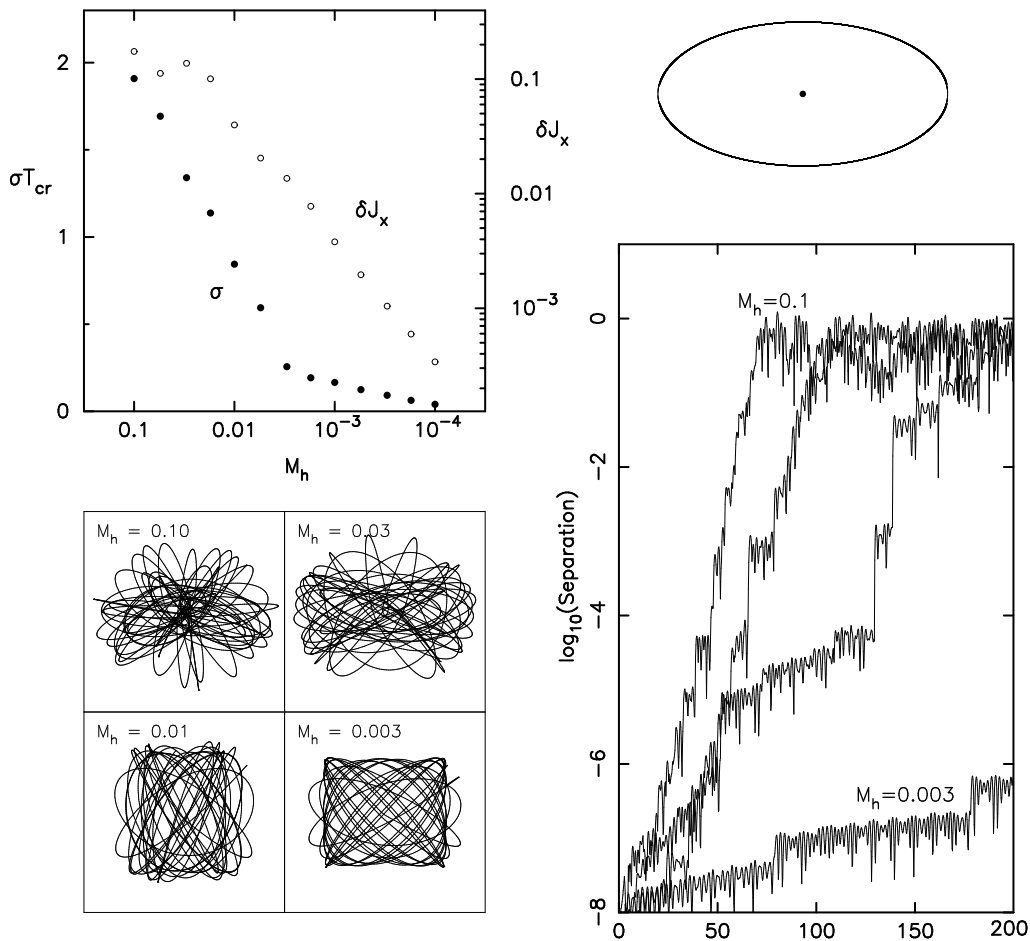}
\caption{Like Fig. 1, for orbits in the potential of a uniform
triaxial ellipsoid of unit mass with an added central point of mass
$M_h$. A single orbit was integrated for each value of $M_h$ and its
properties recorded.  Both the Liapunov exponent $\sigma$ and the
variation in the ``action'' $\delta J_x$ fall off with decreasing
$M_h$.  However, as in Fig. 1, the trajectories can mimic regular
orbits for times much longer than $\sigma^{-1}$ when $M_h$ is small.
The lower right panel again shows why: for small $M_h$, the evolution
of the finite separation between initially adjacent trajectories is
not well predicted by the Liapunov exponent.
\label{fig:BHole}}
\end{figure*}

These results present an interesting contrast with those from Model 2 
(Fig. 2).
Here a single orbit was integrated for each value of $M_h$, the mass
of the central ``black hole,'' in the otherwise smooth,
harmonic-oscillator potential.  
Both $\sigma$ and $\delta J_x$
were now found to vary systematically with perturbation parameter
$M_h$;
thus the orbits tend toward regularity with decreasing $M_h$ as
measured both by their infinitesimal and finite-amplitude behaviors.
But the lower right-hand panel of Fig. 2 reveals essential similarity
with the behavior of the orbits in Model 1.  For large black hole
masses, $M_h\gap 0.02$, divergence continues at roughly the Liapunov
rate until the orbits are separated by of order the size of the
system; while for $M_h\lap 0.01$, the evolution of the separation 
is more complex, with periods of stagnation followed by sudden 
jumps (when the trajectory passes sufficiently close to the 
central point).
Once again, we postulate that the phase space of potentials with
$M_h\gap 0.01$ is globally stochastic, with few impediments to the
motion -- a conjecture that was actually verified (\cite{vm98})
in a different class of triaxial potentials (see \S3).  
For smaller $M_h$, the orbital evolution consists of a sequence of
nearly-random transitions from one, approximately regular orbit to
another (\cite{gb85}); these transitions become rarer as $M_h$ is reduced.

These experiments suggest that Liapunov exponents are not good
predictors of the macroscopic behavior of orbits.
Even trajectories with large $\sigma$ can exhibit nearly regular 
behavior over time scales much longer than $\sigma^{-1}$.  
In deciding whether stochasticity is likely to be important for the 
behavior of orbits, the crucial question is not the value of 
$\sigma$, but whether the exponential divergence {\it
continues} until separations of order the size of the system are
reached.  
If the separation saturates at some value
much smaller than the system size, orbital stochasticity may not
be of much consequence even if $\sigma$ is large.

The sorts of structures that can impede the diffusion of stochastic
orbits are described in more detail in the next section.  Here we note
that even some very simple dynamical models can exhibit the basic
properties that we have described.  For instance, the motion of
particles in a ``Lorentz gas,'' a fixed 2D array of cylindrical
scatters, is generically stochastic, with Liapunov times of order the
time between collisions.  However in the close-packed limit,
trajectories are blocked by the finite size of the scatterers, causing
them to remain confined to narrow regions for long periods of
time (\cite{gas98}).  Thus the exponential instability has little 
consequence for the macroscopic character of the motion.

\section{Phase space structure of triaxial galaxies}

In an integrable potential with $N$ degrees of freedom (DOF), all the
trajectories have exactly $N$ isolating integrals and are confined to
$N$-dimensional tori in phase space.  Such a torus is defined by the
$N$ actions ${\bf J}$ that fix its cross-sectional areas.  Motion
around this torus occurs at a rate determined by a constant frequency
vector ($\omega_1, \omega_2, ..., \omega_N$).  Realistic potentials
with more than 1 DOF are rarely integrable and the motion is more
complex.  While the KAM theorem guarantees that most of the original
tori will persist when an integrable system is slightly perturbed,
even under small perturbations a large part of the phase space will be
influenced by resonances.  A resonant torus is one which satisfies
(one or more) resonant conditions of the form \mbox{\boldmath
$n.\omega$}$ = 0$ between the $N$ fundamental frequencies.  Such tori
are dense in the phase space of the integrable potential but typically
only the lowest order resonances have a significant influence on the
motion in the perturbed potential (\cite{ll92}).  In the vicinity of a
stable resonant torus, motion is still regular, but the orbits have
shapes determined by the order of the resonance -- often very
different from the shapes of orbits in the integrable potential.  In
the vicinity of unstable resonant tori, trajectories are usually
chaotic.

In a system with 2 DOF, the resonant tori are closed periodic orbits
defined by a single frequency $\omega_0=n_2\omega_1=n_1\omega_2$.  In
3 DOF the resonant tori obey a condition on the three fundamental
frequencies of the form $n_1\omega_1+n_2\omega_2+n_3\omega_3 = 0.$
Such a relation does not imply that the orbit is closed, as in 2 DOF,
but rather that it is {\it thin}, densely filling a sheet in
configuration space (\cite{mv99a}).  The resonance condition can be
used to reduce the number of independent frequencies by one; the two
remaining frequencies then describe the rate of rotation around the 2D
reduced torus (\cite{caa98}, \cite{mv99a}).  In 3 DOF systems as in 2
DOF systems, resonant tori are the regions where perturbation
expansions fail, and where the the global structure of phase space is
expected to be changed.  Thus the thin orbits are expected to play a
similar role in 3 dimensions to the role played by periodic orbits in
2 dimensions.

Families of thin orbits may exist even in an integrable potential if
it contains ``primary resonances'' that divide up the phase space
(\cite{ll92}).  This is the case in the well-known ``perfect
ellipsoid'' (\cite{kuz73}, \cite{dez85}): the $1:1$ closed orbits in
the principal planes generate families of thin tube orbits.  In
non-integrable potentials, one expects to find thin orbits and their
associated families throughout phase space.

One of the most useful tools to be employed recently in the study of
galactic potentials is the frequency analysis technique (NAFF)
developed by Laskar (\cite{la90}, \cite{lfc92}, \cite{la98}).
Laskar's technique exploits the fact that regular orbits are
quasiperiodic: Cartesian coordinates like $x(t)$ and $v_x(t)$ can be
expressed as Fourier series in terms of the three fundamental
frequencies on the torus.  While this basic principle has been used to
study orbits in the past (\cite{bs84}), Laskar showed that the
accuracy with which the individual frequency components can be
extracted is greatly improved by employing a Fourier filtering
function and by orthogonalizing successive frequency components.  Once
the entire frequency spectrum is obtained, the three fundamental
frequencies $(\omega_1, \omega_2, \omega_3)$, which appear as linear
combinations in each line, may be identified using an integer
programming algorithm (\cite{vm98}).  All the lines in the spectrum are
then immediately expressible in terms of the fundamental frequencies,
giving a map between the action-angle and Cartesian coordinates
(\cite{vm99b}).

Strictly speaking, only regular orbits are restricted to tori and
amenable to Laskar's technique.  Nonetheless, as discussed above,
stochastic trajectories may mimic regular orbits for long periods of
time.  On short time scales, such an orbit has a frequency spectrum
which mimics a quasi-periodic series.  As the orbit diffuses through
phase space, its frequency spectrum will change.  Laskar (\cite{la93})
showed that over a fixed interval of time ($\Delta T)$ the change in
the frequency of the leading term in the spectrum, $\Delta \omega =
|\omega(T) - \omega(T+\Delta T)|$, is a good measure of the rate of
diffusion of a chaotic orbit in phase space.  Note that Laskar's
$\Delta\omega$, unlike the Liapunov exponent, measures a finite
excursion in phase space.  In this sense it is similar to the $\delta
J$ parameter defined above.

Since Laskar's technique maps the structure of phase space in the
frequency domain, it is ideally suited to identifying the resonant
tori.  In 2 DOF galactic potentials, Papaphillipou \& Laskar
(\cite{pl96}) showed that in a map of $\omega_1/\omega_2$ versus a
third parameter, resonant families appear as regions where
$\omega_1/\omega_2 = $ const over some set of orbits. In 3DOF systems,
a ``frequency map'' may be obtained by plotting the ratios
$\omega_1/\omega_3$ versus $\omega_2/\omega_3$ (\cite{pl98}). Here the
resonant families show up as lines of constant slope.  Thus by
constructing a frequency map of an ensemble of orbits chosen from a
regular grid of starting values, it is possible to locate the
resonances which significantly affect the structure of phase space.

Another useful device is the ``diffusion map'': a plot of the chaotic
diffusion rates $\Delta \omega$ (\cite{la93}).  We calculated
diffusion maps for a family of galaxy models which are triaxial
generalizations of the spherical models described by Dehnen
(\cite{de93}) and others.  These models provide a good fit to the
inner part of the light distributions of real galaxies and have a mass
density given by
\begin{equation}
 \rho(m) = {(3-\gamma) M\over 4\pi abc} m^{-\gamma}
(1+m)^{-(4-\gamma)}, \ \ \ \ 0\le\gamma < 3 
\end{equation}
with
\begin{equation}
m^2={x^2\over a^2} + {y^2\over b^2} + {z^2\over c^2}, \ \ \ \ a\ge
b\ge c\ge 0,
\end{equation}
and $M$ the total mass.  The mass is stratified on ellipsoids with
axis ratios $a:b:c$; the $x$ and $z$ axes are the long and short
axes respectively.  The parameter $\gamma$ determines the slope of the
central, power-law density cusp.

\begin{figure*}
\vspace{14cm}
\includegraphics{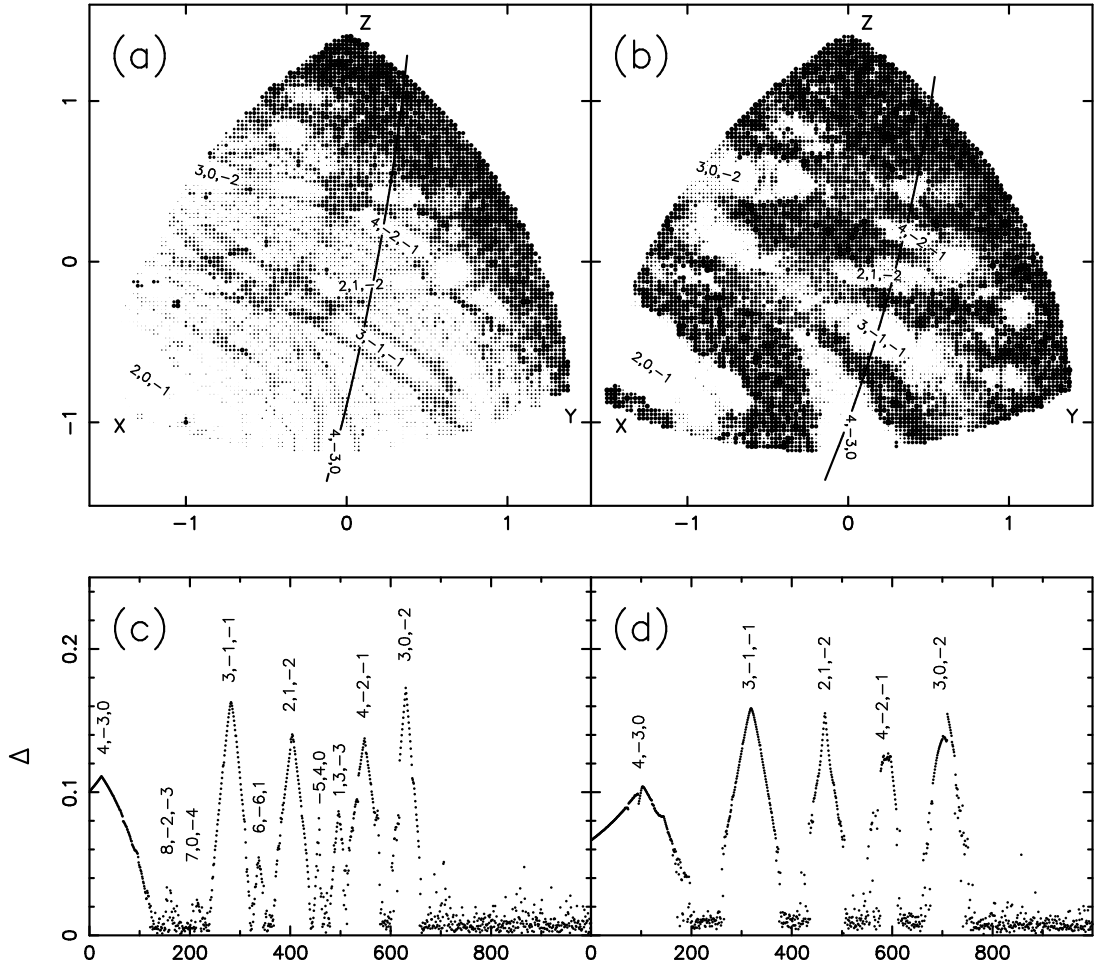}
\caption{Resonances in triaxial potentials. The mass
model in (a) has a weak ($\gamma=0.5$) cusp and no black hole; in (b)
the black hole contains $0.3$\% of the total mass.  Both equipotential
surfaces lie close to the half-mass radius.  The grey scale measures
the degree of stochasticity of orbits started with zero velocity on
the equipotential surface.  Stable resonance zones -- the white bands
in (a) and (b) -- are labeled by the order $(n_1, n_2, n_3)$ of the resonance.
Panels (c) and (d) show the pericenter distance $\Delta$ of a set of
$10^3$ orbits with starting points lying along the heavy solid lines
in (a) and (b).}
%\label{fig:restori}}
\end{figure*}

Figures 3a,b show the diffusion maps at one energy in a triaxial
Dehnen potential with $\gamma=0.5$, with and without a central black
hole.  Orbits were integrated starting at rest on the equipotential
surface just inside the half-mass radius of the model.  The diffusion
rate ($\Delta \omega$) was plotted in grey scale on the equipotential
surface. Rapidly diffusing (chaotic) orbits are coded in black while
regular orbits ($|\Delta \omega| \approx 0$) are shaded white.  The
intensity of the grey scale is proportional to the logarithm of the
stochastic diffusion rate as measured by $|\Delta \omega|$.  The
resonant tori and their associated families appear as narrow white
bands and are marked in the figure by their defining integers $n_1,
n_2, n_3$.  The white bands are flanked by thin dark ones which are
the stochastic layers marking the transitions between resonant
families. Some of these resonances -- the (2, 0 -1) banana; the (4,
-3,0) pretzel; and the (3,0,-2) fish -- are 2D resonances that
correspond to periodic orbits in one of the principal
planes. Miralda-Escud\'e \& Schwarzschild (\cite{mes89}) call such
orbits ``boxlets''. However most of the resonant families cannot be
associated with a periodic orbit.

A striking feature of Fig. 3a is the large number of narrow but
distinct resonance zones. The reason for the narrowness is suggested
by Fig. 3c which shows the distance of closest approach to the
potential center of a set of orbits whose initial conditions lie along
the heavy curve in Fig. 3a. As one passes through a stable resonance,
the orbital pericenter reaches a maximum on the resonance where the
orbit has zero thickness.  The width of the resonance band is set by
how thick an orbit can become before it is rendered stochastic -- this
happens when the distance of closest approach to the destabilizing
center is sufficiently small. In potentials with a steep central
force, this destabilization is attributable to gravitational
deflections which occur when the trajectory passes near the center.
In the case of the $\gamma=0.5$ model of Fig. 3a, the force is
analytic all the way to $r=0$.  Here, the orbits become chaotic
because they come arbitrarily close to an unstable ``separatrix''
layer which marks the transition between one resonant family and the
next.  Once a family is wide enough to pass through the center it can
no longer maintain the same fundamental ``shape'' properties (i.e. it
can no longer be characterized by the same $(n_1, n_2, n_3)$
values). This results in a transition to a new family of orbits. This
transition from one resonant family to another is reminiscent of the
transition from tube to box orbits, which occurs even in integrable
potentials (\cite{dez85}). The lowest-order resonant orbits, have simple
shapes and the largest central ``holes'' allowing a large family
of associated regular orbits.  Higher-order resonances have more
complex shapes and pass nearer the center; their associated families,
and their resonance bands in Fig. 3a, are correspondingly smaller.

Figure 3b shows a diffusion map for a $\gamma = 0.5$ model with a
central black hole of mass $M_h = 0.003 M$.  In this figure the
stochastic regions between the resonant families are significantly
wider.  Figure 3d shows that the black hole destabilizes orbits well
before they become thick enough to pass through the center -- somewhat
different from the standard picture (\cite{gb85}) in which box orbits
are destabilized by passing arbitrarily close to a central black hole.

As the mass of a central point is increased, more and more of the
resonant tori are rendered unstable, and the width of the stochastic
layers and their degree of overlap are increased. Figure rotation has
a similar effect (\cite{va99}). One finds (\cite{vm98}, \cite{mv99a})
that the phase space of box-like orbits (orbits with stationary
points) becomes ``globally chaotic'' when such perturbations are
sufficiently large -- there are essentially no stable invariant tori
left.  For instance, in non-rotating triaxial potentials, global chaos
ensues when the central point contains $\sim 10^{-2}$ of the
galaxy mass within the equipotential surface.  In the globally chaotic
regime, there are few barriers to the motion and one expects orbital
evolution to be very rapid and extensive.  Figure 2 confirms this
prediction in the case of motion in a homogeneous ellipsoid.

\section{Chaotic mixing and dynamical evolution}

\begin{figure*}
\vspace{20.cm}
\includegraphics{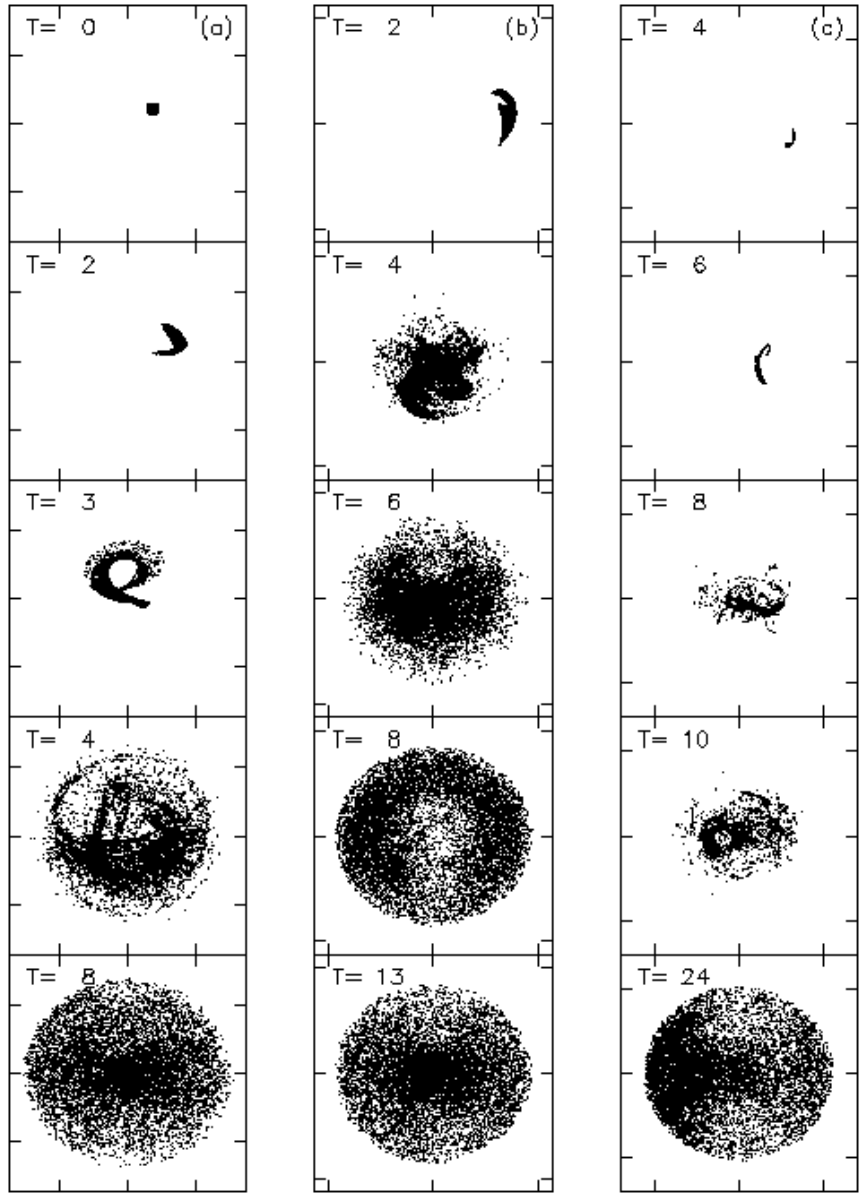}
\caption{Mixing in a triaxial potential with a central point
containing $3\%$ of the total mass.  Time is in units of the local
crossing time.  Ensembles of $10^4$ phase points were distributed
initially ($T=0$) in patches on an equipotential surface with zero
velocity.  Ensemble (a) was begun on a surface enclosing a mass $\sim
3$ times that of the ``black hole''; for ensembles (b) and (c) these
ratios were $\sim 7$ and $\sim 17$ respectively.  Mixing occurs very
rapidly for these ensembles, with a characteristic time of order the
crossing time.
\label{fig:mixing}}
\end{figure*}

As stressed by Kandrup and co-workers (\cite{km94}, \cite{kan90}), a
useful way to think about the consequences of chaotic motion for the
evolution of stellar systems is in terms of mixing.  Mixing is the
process by which an initially non-uniform distribution of points in
phase space relaxes to a uniform one, at least in a coarse-grained
sense (\cite{ll92}).  It therefore provides a link between the
behavior of individual orbits and the evolution of the phase-space
density. While there are many discussions of mixing in the stellar
dynamics literature (\cite{lb67}, \cite{thl86}, \cite{mat88},
\cite{sok96}), most of these are vague with regard the origin of the
mixing or its properties.  Kandrup et al. (1994) and Merritt \& Valluri
(1996) noted many of the special properties of chaotic
mixing. It is irreversible, in the sense that an infinitely fine
tuning is required to undo its effects.  It has an associated time
scale, the Liapunov time, in the sense that a compact set of points
will initially separate at a rate determined by the Liapunov exponent.
Following the arguments presented above, however, we expect mixing to
produce significant changes in the macroscopic distribution of phase
space points only if there are no barriers to the diffusion.

Figure 4 shows examples of chaotic mixing in a triaxial ``imperfect
ellipsoidal'' potential (\cite{mv96}) with a shallow cusp and a
central point mass. Three ensembles of orbits were started at rest on
an equipotential surface and integrated in tandem for several crossing
times. The central point had a mass $M_h = 0.03$ (in units of the
galaxy mass $M$), similar to the masses of the largest black holes
observed in nearby galaxies (\cite{kor95}).  The first ensemble (a) was
begun on an equipotential surface enclosing a mass $\sim 3$ times that
of the ``black hole''; for ensembles (b) and (c) these ratios were
$\sim 7$ and $\sim 17$ respectively. The number in each frame is the
elapsed time in units of the local crossing time. All ensembles were
initially distributed in a tiny square patch as for $T=0$ in ensemble
(a).

Mixing occurs very rapidly in these ensembles.  At the lowest energy,
(ensemble a), the linear extent of the points in configuration space
doubles roughly every crossing time until $T\approx 4$, when the
volume defined by the equipotential surface appears to be nearly
filled.  At the highest energy (ensemble c), mixing is slower but
substantial changes still take place in a few crossing times.  The
final distribution of points at this energy  still shows some structure,
reminiscent of a box orbit.

Mixing in triaxial potentials with smaller central masses 
can be slower, requiring hundreds or even thousands of crossing 
times to reach a near-invariant state (\cite{mv96}) -- even though 
the Liapunov exponents in these potentials are comparable
(expressed in units of the local orbital frequency) to 
those of the orbits in Fig. 4.  Here again, we find that Liapunov 
exponents are poor predictors of the rate of evolution.

The mixing illustrated in Fig. 4 takes place in a phase space that is
globally chaotic, with almost no regular orbits.  In triaxial
potentials containing a central black hole, a ``zone of chaos''
extends from the nucleus out to a radius containing $\sim 10^2$ times
the mass of the central point (\cite{mv99a}). Each of the ensembles in
Fig. 4 lies within this zone.  Near the outer edge of the chaotic zone
(ensemble c), the mixing is beginning to become non-random, and the
distribution of points at the final time step still shows hints of
structure. At still larger radii, or in potentials with smaller black
holes, phase space is a complicated mixture of regular and chaotic
trajectories (cf. Fig. 3) and mixing is correspondingly slower.

Mixing like that of Fig. 4 should produce rapid changes in the shape
of a galaxy. Such evolution has in fact been observed in $N$-body
simulations of the response of a triaxial galaxy to the growth of a
central black hole. Merritt \& Quinlan (\cite{mq98}) found that a
triaxial galaxy evolves to axisymmetry in little more than the local
crossing time at each radius when the black hole mass exceeds $\sim
2.5\%$ of the total galaxy mass. There are two factors that can cause
mixing at large radii to occur faster in these $N$-body simulations
than in Fig. 4(c). First, in a self-consistent potential all orbits
are initially well spread out and not confined to tiny patches as in
the mixing experiments. Second, mixing timescales at small radii are
almost 100 times faster (in absolute time units) than at the half mass
radius. Once the inner region changes shape the orbits further out are
no longer in equilibrium in the new potential and can respond more
rapidly than they would in a fixed potential (\cite{bar99}).
Therefore the orbits at larger radii respond collectively both to the
perturbing central black hole and to the potential changes at small
radii. Smaller black holes induce slower evolution, consistent (at
least qualitatively) with the less efficient mixing expected in such
potentials (\cite{mv96}).  Such experiments -- which should be
repeated with a wider range of initial conditions, including rotating
models -- confirm that chaos can be an important mechanism in
determining the global structure of galaxies.

\section{Conclusions}

The orbits of stars in stellar systems are generically chaotic. 
Exponential sensitivity to perturbations implies that initially 
adjacent trajectories will diverge in a fraction of a crossing time. 
However this divergence does not necessarily imply a macroscopic 
evolution of the phase space distribution on a comparable time 
scale, as has been suggested by some authors.
The reason is that perturbations may grow exponentially only for a limited 
time, saturating when separations are much smaller than the size 
of the system.  This occurs, for instance, in a potential composed of 
$N$ point masses when $N$ is large. In such potentials, 
orbits can mimic regular 
orbits for many oscillations even though their Liapunov exponents are 
large. Only when perturbations are able to grow until their scale is of 
order the size of the system does the chaos have a significant 
influence on the macroscopic dynamics.

The redistribution of stars in phase space due to chaos may be 
simulated by mixing experiments in fixed potentials.  Mixing
leads to a near-invariant distribution of points within the 
accessible phase-space region.  We find that mixing 
in a globally chaotic region of phase space -- a region where 
almost all the orbits are stochastic -- is very efficient, producing 
substantial changes in the distribution of points in just a 
crossing time.  In phase space regions containing a mixture of 
regular and chaotic trajectories, mixing can be slower.
The objects that impede mixing appear to be the resonant tori 
and their associated families of regular orbits. In triaxial 
potentials containing a central point mass, mixing becomes very 
efficient when the central mass is big enough to destroy all of 
the resonant tori.  The result is large-scale evolution of the 
galaxy toward axisymmetric shapes.

\end{document}